\documentclass[aps,pra,twocolumn,superscriptaddress,nofootinbib]{revtex4-2}
\usepackage{amssymb,graphicx,color}
\usepackage[intlimits]{amsmath}
\usepackage[english]{babel}
\usepackage[colorlinks]{hyperref}
\hypersetup{
colorlinks=blue,
linkcolor=blue,
filecolor=blue,
urlcolor=blue,
citecolor=blue
}
\pdfoutput=1

\usepackage[normalem]{ulem}
\usepackage{comment}
\usepackage{cprotect}
\usepackage{ragged2e}
\usepackage{enumitem}
\usepackage{amsmath}
\usepackage{mathtools}
\usepackage{amsmath}
\usepackage{amsfonts}
\usepackage{amssymb}
\usepackage{siunitx}
\usepackage{verbatim}
\usepackage{hyperref} 
\usepackage{natbib} 
\usepackage{amsthm}
\usepackage{amsfonts}
\usepackage{braket}
\usepackage{float}
\usepackage{blkarray}
\usepackage{physics}
\usepackage{qcircuit}
\usepackage{multirow}
\usepackage{tikz}
\usetikzlibrary{calc}
\usetikzlibrary{matrix}
\begin{document}

\preprint{APS/123-QED}

\title{Photonic Simulation of Localization Phenomena Using Boson Sampling}

\author{Anuprita V. Kulkarni}
\affiliation{Department of Physics, Indian Institute of Science Education and Research, Bhopal 462066, India}

\author{Vatsana Tiwari}
\affiliation{Department of Physics, Indian Institute of Science Education and Research, Bhopal 462066, India}

\author{Auditya Sharma}
\affiliation{Department of Physics, Indian Institute of Science Education and Research, Bhopal 462066, India}

\author{Ankur Raina}
\email{ankur@iiserb.ac.in}
\affiliation{Department of Electrical Engineering and Computer Science, Indian Institute of Science Education and Research, Bhopal 462066, India}

\begin{abstract}
Quantum simulation in its current state faces experimental overhead in terms of physical space and cooling. We propose boson sampling as an alternative compact synthetic platform performing at room temperature.
Identifying the capability of estimating matrix permanents, we explore the applicability of boson sampling for tackling the dynamics of quantum systems without having access to information about the full state vector. By mapping the time-evolution unitary of a Hamiltonian onto an interferometer via continuous-variable gate decompositions, we present proof-of-principle results of localization characteristics of a single particle. We study the dynamics of one-dimensional tight-binding systems in the clean and quasiperiodic-disordered limits to observe Bloch oscillations and dynamical localization, and the delocalization-to-localization phase transition in the Aubry-Andr\'e-Harper model respectively. Our computational results obtained using boson sampling are in complete agreement with the dynamical and static results of non-interacting tight-binding systems obtained using conventional numerical calculations.
Additionally, our study highlights the role of number of sampling measurements or ``shots" for simulation accuracy.

\end{abstract}

\maketitle


\section{\label{sec:introduction}Introduction}

The Noisy Intermediate-Scale Quantum (NISQ) era\cite{preskill2012quantum_nisq, Preskill_2018_nisq, mbqm_nisq}, characterized by near-term quantum processors of a limited number of qubits ($<1000$) without full fault-tolerance, has popularized the idea to use linear interferometers \cite{resurgenceoflinearoptics} - especially linear optics interferometers - as a processor of quantum information. For example, the boson sampling problem involves sampling from the output distribution of $N$ indistinguishable bosons evolving through a passive $M\times M$ linear interferometer defined by a Haar-random unitary matrix. Aaronson and Arkhipov\cite{AABS, bs_review} showed that boson sampling involves the computation of permanents of complex matrices, which is at least \#P-hard, widely believed to be classically intractible\cite{VALIANT1979189, permanent_polytimeapprox}. Initially, boson sampling was viewed only as a candidate for demonstration of quantum computational advantage\cite{bsonaphotonicchip, integrated_interf_photonic_bs, experimental_bs, pbs_tunablecircuit, bs_gaussianstate_llrk, bs_assessment, quantumoptics_complexitytheory, multiphoton_interf, bs_review, class_sim_QO, scaling_bs_exp, bs_singlephotonfock, classical_bs_superiornearterm, higheff_multiphoton_bs, quantumsamplingprobs}. Recent works have utilized boson sampling in a variety of avenues like quantum random number generators\cite{unbiased}, cryptography\cite{bs_crypt1,bs_crypt3, photonicdatalocking, bs_crypt2},  image encryption\cite{image_encrypt}, and parametrized quantum circuit learning\cite{parametrized_bs}. A boson sampling variant using Gaussian states as input\cite{bs_gaussianstate_llrk, gbs, gbs_detail} has been applied towards computation of molecular vibronic spectra\cite{gbs_vibronic} and drug discovery\cite{gbs_drugdiscovery}. Our work adds a potential application of boson sampling in the direction of quantum simulation, which is a useful step towards tackling many body problems.

Quantum simulation\cite{qsim3, qsim2, qsim1, cv_qho, topologicalquantumwalksonnisq} facilitates the study of single and many-body physics by mapping complex Hamiltonians to alternative quantum systems, enabling the exploration of challenging experimental cases and mitigating computational difficulties from growing Hilbert-space dimensions\cite{scalable}. Boson sampling circuits have been used to study quantum systems through simulation of spin Hamiltonians\cite{quantumsim_bscircuit}, to probe quantum signatures of chaos from the perspective of Floquet theory\cite{bastidas2023}, and in quantum chemistry studies of molecular electronic structures\cite{bs_quantumchem}. With the motivation of simulating quantum dynamics, we successfully adopt boson sampling to study for the first time the transport properties and localization in tight-binding lattices which host non-equilibrium features in both the clean and disordered limits.

We realize a photonic simulation of the non-interacting one-dimensional tight-binding model\cite{localization1981, localization_2019, quantumtransport_npj, characteristic_lengthscale}, which in the clean limit subjected to static and time-periodic electric field is well known to exhibit the phenomena of Bloch oscillations\cite{bloch1929, zener1934, dynamicsofbloch, blochosc_expobs_1d2d, blochosc_expobs_nature} and dynamical localization\cite{dynamicallocalization, dunlapkenkre, dynamicallocwithbec, weld2024} at special drive-parameters, respectively. Although well-understood theoretically, experimental observation of these phenomena is difficult owing to short coherence times and lattice imperfections causing scattering rates to be faster than the expected time period of the oscillations\cite{characteristic_lengthscale, blochosc_expobs_nature}. We are able to clearly observe these phenomena on the simulated photonic circuit.
Next, we also study a quasiperiodic disordered model popularly known as the Aubry-Andr\'e-Harper model\cite{Harper1955, aubry_1980_analyticity} which is known to host the transition from extended to localized phase on varying the quasiperiodic disorder strength\cite{aubry_1980_analyticity, aahattention3, aahattention4, aahattention5, aahattention6, aahattention7, aahattention8, aahattention1, aah1, aah_spectral_corr, aahproperties_pwave, weld2024}.

\begin{figure*}
    \centering
    \includegraphics[width=1.02\linewidth]{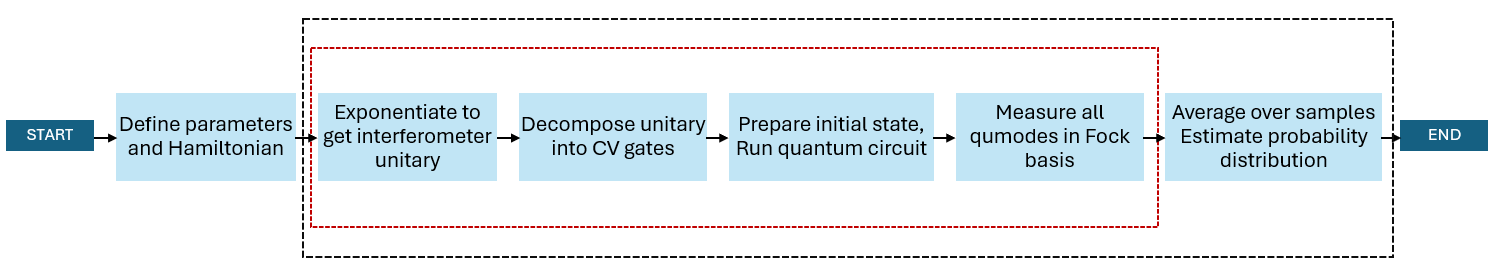}
    \cprotect\caption{An illustration of the methodology to study quantum dyamics using boson sampling. The steps enclosed within the black box denote the actions to be implemented at each timestep to be considered. The steps enclosed within the red box may be performed for a high number of `shots' for an increasingly accurate reconstruction of the probability distribution upon averaging over the samples.}
    \label{fig:flowchart}
\end{figure*}
We depict the methodology for using boson sampling as a platform for quantum simulation in Fig. (\ref{fig:flowchart})~\cite{hamsim_linearcombuniop,hamsimalgo}. Upon defining the Hamiltonian of the system of interest, an input state is prepared and passed into the circuit which characterizes its unitary evolution, followed by a photon-counting measurement across all modes. Sampling from this circuit multiple times allows us to reconstruct the probability distribution, estimating permanents of unitary sub-matrices in the many-body case. Thus, without having access to the full state information and using only the estimated probability distribution, we show that it is possible to compute the dynamics of closely related quantities such as mean-squared displacement, participation entropy and inverse participation ratio. Our results recover the signature of Bloch oscillations in time-independent systems and dynamical localization and delocalization in time-periodic systems. We also capture the delocalization-to-localization phase transition of the Aubry-Andr\'e-Harper model with the help of the study of the dynamics of participation entropy and inverse participation ratio. Thus, our work establishes an excellent concurrence between the dynamics obtained by boson sampling and those obtained via numerical computations.

We  therefore propose boson sampling on some integrated photonic chip\cite{bsonaphotonicchip, nanophotonic_IC, nanophotonic_chip} as an alternative candidate to probe these dynamics. Photonic circuits which can operate at room temperature\cite{roomtemp} enable a simpler way to study quantum dynamics of systems with greater experimental overheads.

The remainder of this work is organized as follows. Section \ref{sec:theory} gives a theoretical overview of boson sampling and the mathematical formalism behind the mapping of a Hamiltonian to a photonic circuit. Section \ref{sec:elec} covers the model Hamiltonian and results of the dynamics of a tight-binding chain in the presence of an electric field using the boson sampling method. 
Section \ref{sec:aah} presents the Aubry-Andr\'e-Harper model and explores the delocalization to localization phase transition through dynamics of participation entropies and inverse participation ratios, followed by a discussion and future outlook in Section \ref{discussion}.

\section{\label{sec:theory}Theory}

This section provides a brief introduction to the theoretical formalism of the boson sampling method relevant to this work.
\subsection{Theoretical Overview of Boson Sampling}
\begin{figure}[H]
    \centering
    \includegraphics[width=\linewidth]{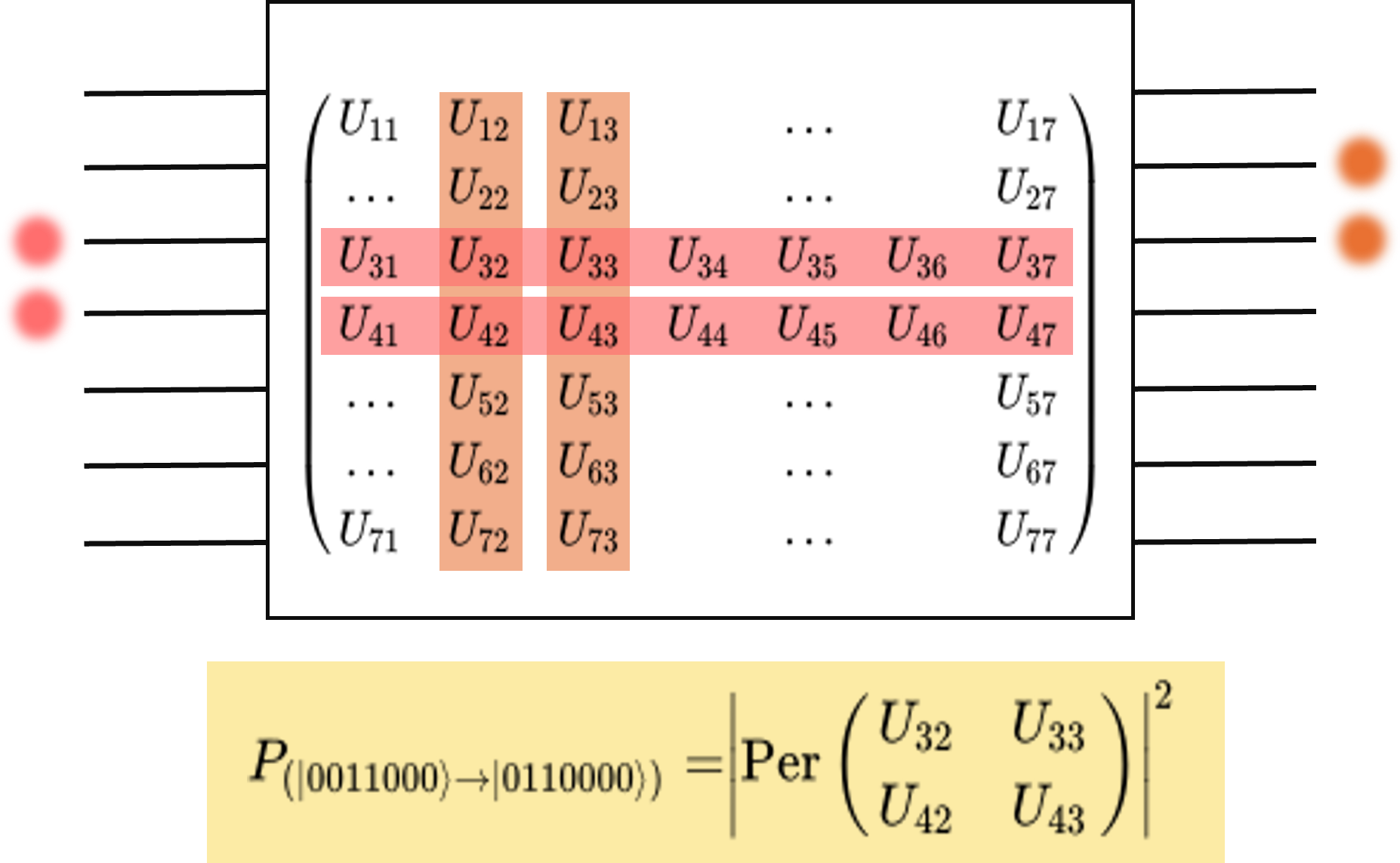}
    \caption{An example of a boson sampling circuit with $M=7, N=2$. The probability of finding the photons (Eq. (\ref{perm})) in a particular configuration depends on the permanent of a $2\times2$ sub-matrix of the $7\times7$ unitary matrix characterizing the interferometer. The columns of the sub-matrix are determined based on the output state, and the rows of the sub-matrix based on the input state.}
    \label{fig:bs_schematic}
\end{figure}

We consider a circuit with $M$ modes, each having an associated pair of bosonic creation
and annihilation operators\cite{bs_review, quantumsamplingprobs}. An initial state in the Fock basis of $N$ photons can be written as 
\begin{equation}
    \ket{\psi} = \ket{n_1^{(I)},n_2^{(I)},...,n_M^{(I)}} = \prod_{j=1}^M \frac{(\hat{a}_j^\dag)^{n_j^{(I)}}}{[n_j^{(I)}!]^{1/2}} \ket{\textbf{0}},
\label{init_state}
\end{equation}
where $\ket{\textbf{0}} = \ket{0_1, 0_2, ... , 0_M}$ represents the vacuum state.
The above equation specifies the initial configuration $I$ of $N$ particles distributed across $M$ modes, with $n_j^{(I)}$ specifying the photon number in a particular mode indexed by $j$. 
The initial state $\ket{\psi}$ is evolved via a passive linear optics interferometer, which implements a unitary map on the creation operators 
\begin{equation}
    \hat{U}\hat{a}_i^\dag \hat{U^\dag} = \sum_{j=1}^{M} U_{i,j}\hat{a}_j^\dag,
\label{unitary_transformation}
\end{equation}
where $\hat{U}$ is a unitary matrix characterizing the linear optics network. The output state $\ket{\psi'}$ thus obtained is a superposition state of all the various possible configurations\cite{bastidas2023}:
\begin{equation}
    \begin{aligned}
        \ket{\psi'} &= \prod_{i=1}^{M} \frac{[\hat{U}\hat{a}_i^\dag \hat{U}^\dag]^{n_i^{(I)}}}{[n_i^{(I)}!]^{1/2}} \ket{\textbf{0}} \\
                 &=\prod_{i=1}^{M} \frac{[\sum_{j=1}^{M} U_{i,j}\hat{a}_j^\dag]^{n_i^{(I)}}}{[n_i^{(I)}!]^{1/2}} \ket{\textbf{0}} \\
                 &= \sum_F \gamma_F\ket{n_1^{(F)},n_2^{(F)},...,n_M^{(F)}},
    \end{aligned}\label{finalket}
\end{equation}
where $F$ indexes all the possible configurations of the final state, $n_i^{(F)}$ represents the number of photons in the $i^{th}$ mode for configuration $S$, and $\gamma_F$ is the amplitude associated with the configuration $F$. The probability $P_F$ of obtaining a particular configuration $F$ will be given by $P_F = |\gamma_F|^2$. 
The probability amplitudes of \(N\) particle boson sampling are given by permanents of \(N \times N\) sub-matrices of the \( M \times M \) interferometer unitary. These sub-matrices are constructed with rows chosen according to the input state and columns chosen according to the output state, with repeated copies of the corresponding row/column in case of coincidence events. If $U^{(F, I)}$ denotes the sub-matrix for a specific configuration, the probability of obtaining that configuration is given by
\begin{equation}
P_F = \lvert \gamma_F\rvert^2 = \frac{\lvert \mathrm{Per}[U^{(F,I)}]\rvert^2}{\prod_{i = 1}^{M}n_i^{(I)}!\prod_{i=1}^{M}n_i^{(F)}!}. \label{perm}
\end{equation}
An example of a boson sampling circuit and the relation between probabilities and permanents (Eq. (\ref{perm})) is displayed in Fig. (\ref{fig:bs_schematic}).

\begin{figure*}
    \centering
    \includegraphics[width=0.85\linewidth]{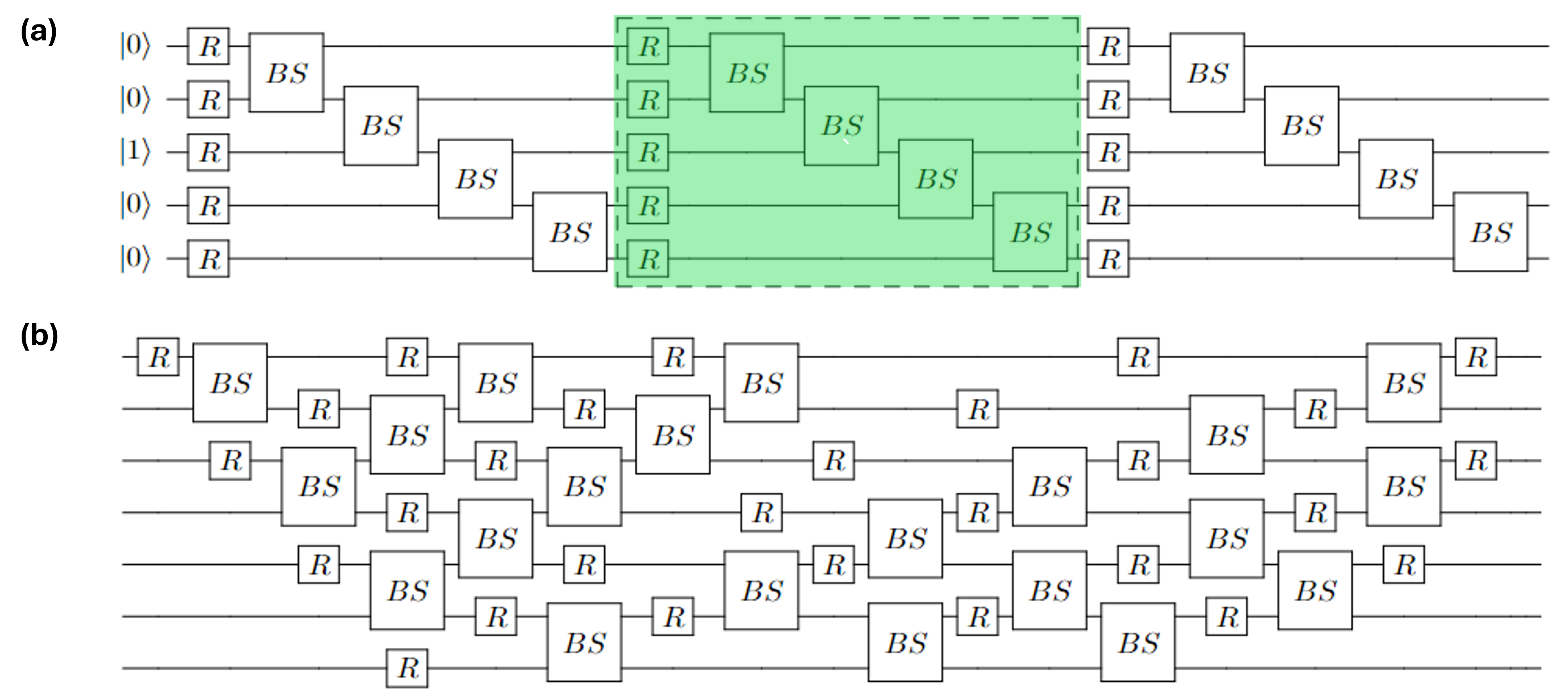}
    \cprotect\caption{ \textbf{(a)} An example of the decomposition of a time-evolution unitary of an arbitrary Hamiltonian (Eq. (\ref{eq:arbitraryhamiltonian})) into continuous-variable quantum gates (Eq. (\ref{unitarytransformation})). The system size here is $M = 5$ and a single photon ($N=1$), with $l=3$ as per Eq. (\ref{unitarytransformation}). \textbf{(b)} An example of the boson sampling circuit ($M=7$) for simulation the dynamics of the static case of Eq. (\ref{hamiltonian_bloch}) upto the first timestep. This interferometer has been characterized according to a rectangular decomposition\cite{clements} with $M(M-1)/2 = 21$ beam-splitters using the \verb|strawberryfields.ops.Interferometer| class.}
    \label{fig:examplecircuit}
\end{figure*}
We can connect boson sampling to the time evolution of a tight-binding chain by defining the unitary transformation on the creation and annihilation operators according to the propagator of the desired Hamiltonian \cite{gopi}. To effectively simulate quantum dynamics on this platform, it is crucial to determine the appropriate interferometer decomposition equivalent to the transformation described by the propagator in question\cite{recketal, clements}.

\subsection{\label{subsec:mapping}Mapping: Hamiltonian to Photonic Circuit}

In this section, we present how we map a Hamiltonian of interest to a photonic circuit such that a run of the circuit is equivalent to the time-evolution of a quantum system governed by the desired Hamiltonian, upto a given point in time\cite{hamsim_linearcombuniop, hamsimalgo}. It should be noted that running such a circuit does not directly give one access to the state of the system following the evolution, however multiple photon-counting measurements can help construct the probability distribution of the photon, to be used for further related computations.

We consider an arbitrary tight-binding Hamiltonian
\begin{equation}
    \hat{H} = J\sum_{j}\sum_{k} (A_{jk}\hat{a}_j^\dag\hat{a}_k + \text{h.c}) + \sum_{j}h_j\hat{a}_j^\dag\hat{a}_j.\label{eq:arbitraryhamiltonian}
\end{equation}
The time-evolution operator corresponding to this Hamiltonian is 
\begin{equation}
    \hat{U}(t_0, t) = \exp(-i \int_{t_0}^t H(s)ds),
\end{equation}
which reduces in the time-independent case to 
\begin{equation}
    \hat{U}(t_0, t) = \exp(-i H(t-t_0)).\label{unitarytimeevol}
\end{equation}
Without loss of generality, consider $t_0 = 0$. Using the Lie-Product Formula\cite{lietrotter, hamsim_linearcombuniop} and the bosonic commutation relation $([\hat{a}_j^\dag, \hat{a}_k] = \delta_{jk} \forall (j,k)\implies [\hat{n}_j, \hat{n}_k] = 0)$ for our Hamiltonian, we get:
\begin{equation}
    \begin{aligned}
\hat{U}(t) 
&= \lim_{l \to \infty}\Biggl[\exp{\frac{-i Jt}{l}\sum_{j}\sum_{k} (A_{jk}\hat{a}_j^\dag\hat{a}_k + \text{h.c})}\\
&\phantom{=} \prod_j\exp\Bigl(-\frac{i t}{l}h_j\hat{n_j} \Bigr)\Biggr]^l + O(t^2/l).\label{unitarytransformation}
    \end{aligned}
\end{equation}
In our case, a beam-splitter between two consecutive modes characterizes the operator implementing the exponential of a nearest-neighbor hopping term, while a phase-space rotation accounts for the exponential of the onsite term. The key information about the Hamiltonian such as $J, A_{ij}, h_i , t$ can be encoded into the beam-splitters and phase-shifters as\cite{cv_gatedecompositions1, cv_gatedecompositions2}
\begin{align}
    \theta = -\frac{Jt}{l}\label{eq:theta},\\
\phi_{jk} = -i\log{(A_{jk})}\label{eq:phi},\\
\delta_j = -\frac{t}{l}h_j\label{eq:delta}.
\end{align}
Therefore, the time evolution of any non-interacting bosonic system can be encoded into a linear optical network. At a single particle level, a mapping to the linear optics network can admit both bosonic and fermionic systems, however only bosonic systems may be simulated in the many-body case. Additionally, since boson sampling is characterized by a passive linear optics network (Eq. \ref{unitarytransformation}), the unitary transformations admissible to a boson sampling problem must be at most quadratic in the creation and annihilation operators i.e. the Hamiltonians must be those of non-interacting systems only\cite{bastidas2023}.

\subsection{Methodology for Simulation of Dynamics}\label{subsec:methodology}
All boson sampling simulations in this work are implemented using Xanadu's Strawberry Fields\cite{sfpaper, sfpaper2} library for photonic quantum computation. We use the syntax of the Blackbird quantum assembly language, which can characterize the fundamental operations of state preparation, gate application and measurements on continuous-variable systems. Each quantum circuit consists of `qumodes' which act as the input wires to an interferometer. 
We explore the single-particle scenario, wherein a Fock state $\ket{1}$ is prepared at a desired qumode with the rest of the qumodes defaulting to the vacuum state $\ket{0}$. Next, for a particular timestep, an instance of a boson sampling circuit is created, which is characterized by the unitary operator specifying the time-evolution of the desired Hamiltonian.

This unitary operator is decomposed onto an interferometer in the form of beam-splitters and phase-shifters as detailed in Eqs. (\ref{unitarytransformation}), (\ref{eq:theta}), (\ref{eq:phi}), and (\ref{eq:delta}). Upon running this circuit for a desired number of `shots', a set of samples is obtained, which can be averaged over to estimate the probability distribution. Contrary to the many-body case where boson sampling simulations will estimate permanents of $N\times N$ unitary sub-matrices (for $N$ particles), this problem reduces to recovering a single matrix element in the single particle case. This process is repeated for every timestep in the evolution of the system. An illustration of this protocol can be found in Fig. \ref{fig:flowchart}. 
In Fig. \ref{fig:examplecircuit} (a), we illustrate a basic decomposition of a time-evolution unitary into beam-splitters and phase-shifters, as per Eq. (\ref{unitarytransformation}). It is possible to characterize the interferometer unitary through various methods of decomposition\cite{recketal, clements}. The accuracy of a particular decompositoin can be measured via the gate fidelity between a particular $M\times M$ unitary $\hat{U}$ and its decomposed counterpart $\hat{U}_D$, given as\cite{fidelity} 
\begin{equation}
\text{Fidelity} = |\text{Tr}(\hat{U}\hat{U}_D^\dag)|/M.
\end{equation}
The boson sampling results presented in this work use a method from the Strawberry Fields library that returns all Fock basis probabilities at the end of the evolution. For relatively small system sizes, this is a convenient method; however, it does not seem sustainable, considering the overall complexity of the problem. Additionally, in a real experiment, one would not have direct access to Fock probabilities and hence the accuracy of the results will depend on the number of samples taken at each time-step. The effect of this is illustrated in the \hyperref[appendix:samples]{Appendix}.

\section{\label{sec:elec}Electric Field Driven Tight-Binding Chain}

The first model we investigate with the help of boson sampling is a one-dimensional tight-binding system subjected to an external electric field.  In case of a uniform static electric field, a charged particle undergoes oscillatory motion known as Bloch oscillations~\cite{bloch1929, zener1934,  electronsinuniformfield, dynamicsofbloch, blochosc_expobs_nature, noise_blochoscandwsl}. The presence of time-periodic electric field yields the phenomenon of dynamic localization of the system on adequate tuning of the driving amplitude and frequency\cite{dynamicallocalization, dunlapkenkre, dynamicallocwithbec, dynamicalloc_quasiperiodic}. The Hamiltonian for this model is
\begin{equation}
    \hat{H} = -J\sum_{i=0}^{L-1}(\hat{c}_{i+1}^\dag\hat{c}_i + \text{h.c.}) + aF(t)\sum_{i=0}^{L}\Bigl(i - \frac{1}{2}\Bigr)\hat{c}_i^\dag\hat{c}_i,
    \label{hamiltonian_bloch}
\end{equation}

where $\hat{c^\dag}$ and $\hat{c}$ are single-particle creation and annihilation operators respectively, $J$ is the strength of the hopping,  $a$ is the lattice parameter (which we shall set equal to 1), and $F(t)$ is the electric field.
\begin{figure}
    \centering
    \includegraphics[width=\linewidth]{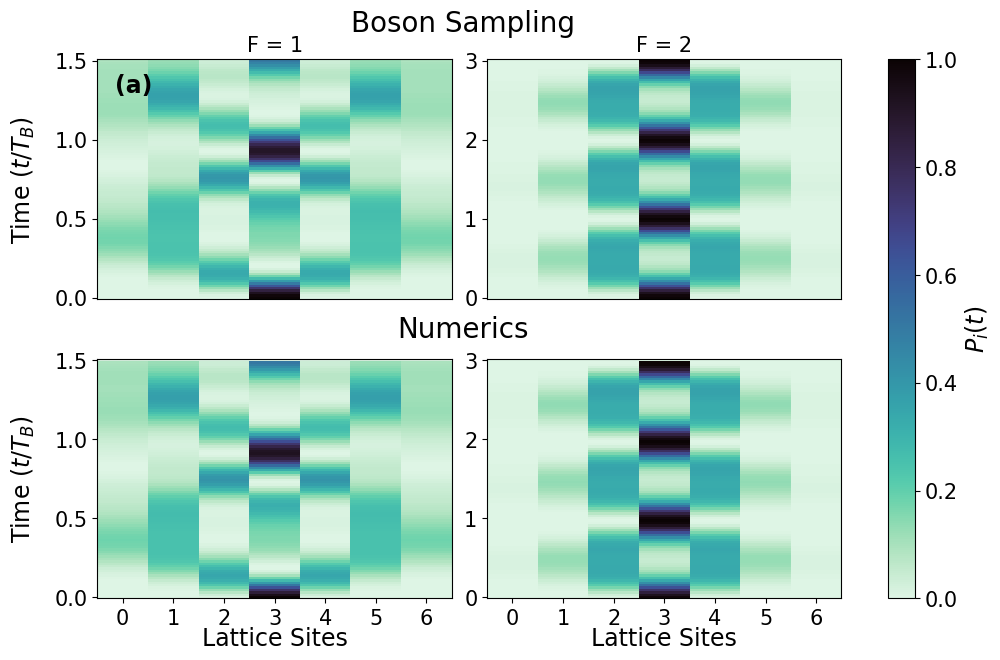}
    \caption{\textbf{Static Field (a):} Evolution of onsite populations $P_i(t)$ in the presence of a static electric field, obtained via boson sampling (first row) and numerical computations (second row). The $F=1$ case illustrates boundary effects in a device of size smaller than the Bloch length, whereas Bloch oscillations are clearly visible in the $F=2$ case.}
    \label{fig:static_heatmap}
\end{figure}
\begin{figure}
    \centering
    \includegraphics[width=0.8\linewidth]{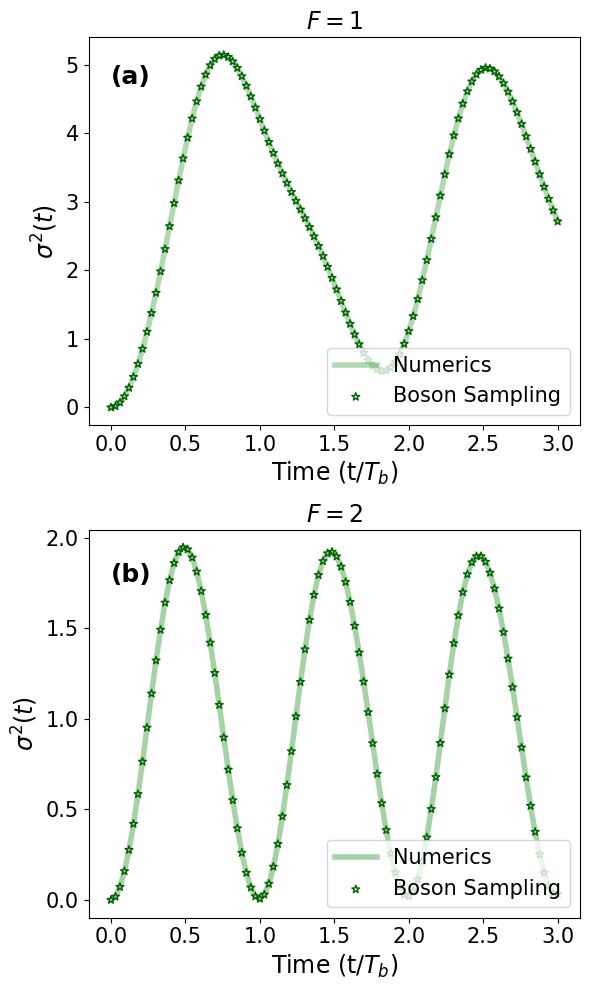}
    \caption{\textbf{Dynamics of the system subjected to static electric field (a), (b):} Comparison of results between numerics and boson sampling for mean squared displacement as a function of time for field strengths $F = 1$ and $F = 2$ respectively. Parameters considered are $M = 7$, $J= 1$, $t = 3T_b$.}
    \label{fig:msd_bloch}
\end{figure}
\begin{figure*}
    \centering
    \includegraphics[width=0.47\linewidth]{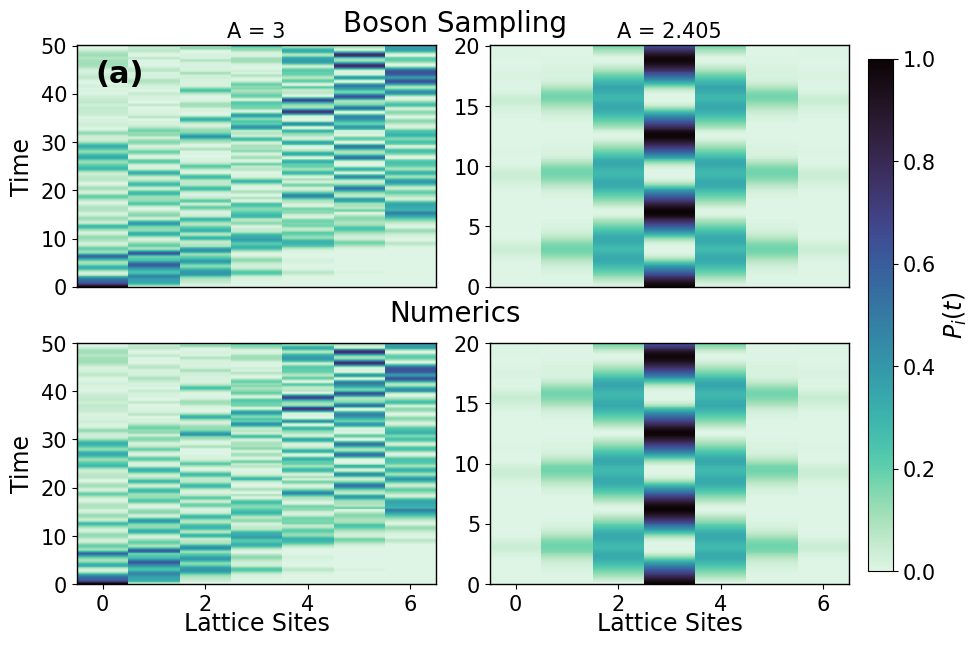}
    \includegraphics[width=0.35\linewidth]{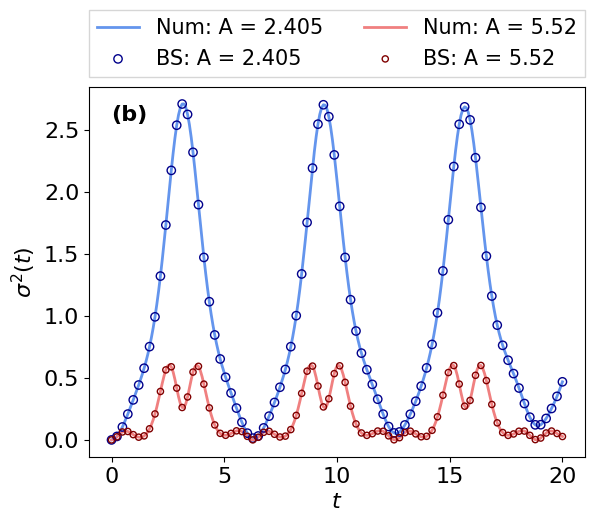}
    \caption{\textbf{Dynamics of a time-periodically driven system.} \textbf{(a)} The time-evolution of onsite populations $P_i(t)$ for a time-periodically driven system obtained via boson sampling (first row) and numerical computations (second row). For arbitrary ratio $A/\omega = 3$, the system is delocalized. However, when $A/\omega$ has been adjusted to be a root of the zeroth order Bessel function ($A/\omega = 2.405$), dynamical localization is observed. The initial state is $\ket{1000000}$ for the $A=3$ case, and $\ket{0001000}$ for the $A=2.405$ case. Parameters considered here are $J = 0.5$, $\omega=1$. \textbf{(b)} Dynamics of mean-squared-displacement $\sigma^2(t)$ at the dynamical localization points $(A=2.405\omega, 5.52\omega)$. We can see a slight drift in the plot at later times - this erroneous behaviour can be attributed to the choice of a larger step-size for computing this evolution. The system size considered is $M = 7$, with parameters $J = 0.5$, $\omega = 1$, $\hbar = 1$}
    \label{fig:dl}
\end{figure*}

We consider two cases of the applied electric field:\\
\begin{equation}
    F(t) = \begin{cases}
F & \text{static field}\\
A\cos{(\omega t)} & \text{sinusoidal driving}
\end{cases}.\label{eq:cases}
\end{equation}
The energy spectrum of the static Hamiltonian given by Eq. (\ref{hamiltonian_bloch}, \ref{eq:cases}) forms the well-known equidistant Wannier-Stark ladder $E_m = maF$  ($m = 0, \pm 1, \pm 2,...$), with eigenstates given by Wannier-Stark states
\begin{equation}
    \ket{\psi_m} = \sum_l \mathcal{J}_{l-m}(\gamma)\ket{m}, \label{eq:wannierstates}
\end{equation}
where $\gamma=2J/aF$, $\mathcal{J}_{l-m}$ is a Bessel function known to be mainly localized in the interval $|l - m| < \gamma$\cite{dynamicsofbloch, localization_in_ac}. The Wannier-Stark states are thus extended over the interval $\gamma a = 2J/F$. This determines the length-scale of Bloch oscillations, i.e. the Bloch length $L_B$ defined as\cite{characteristic_lengthscale}
\begin{equation}
    L_B = \frac{2J}{F}.\label{eq:blochlength}
\end{equation}
The time period of these oscillations, i.e. the Bloch period $T_B$ is
\begin{equation}
    T_B = \frac{2\pi\hbar}{aF}
 \implies \omega_B = \frac{aF}{\hbar}.
\end{equation}
On the other hand, a time-periodic sinusoidal driving yields dynamical localization\cite{dunlapkenkre, dynamicallocwithbec} upon adequate tuning of the amplitude and frequency of the drive. The theory of dynamical localization has been discussed in detail in Ref. \cite{characteristic_lengthscale, dunlapkenkre}; we nonetheless include a brief summary.
The solutions to the time-dependent Schr\"{o}dinger's equation of the effective Hamiltonian for the sinusoidal drive (Eq. \ref{eq:cases}) are given by the accelerated Bloch states or Houston states\cite{houston} as
\begin{equation}
    \ket{\psi_k(t)} = \exp{-\frac{\iota}{\hbar}\int_0^t\dd \tau E(q_k(\tau))} \sum_l e^{\iota q_k(t)l}\ket{l},\label{houston}
\end{equation}
where $\ket{l}$ denotes the Wannier state localized at the $l^{th}$ lattice site. The dispersion relation $E(q_k(t))$ is a modified version of that of the nearest-neighbor tight-binding model $E(k) = -2J\cos(ka)$, with a time-dependent quasi-momentum $q_k(t) = k + \frac{A}{\hbar\omega}\sin(\omega t)$.  We can decompose the Houston states in terms of Floquet states with quasi-energies $\epsilon(k)$\cite{dynamicallocalization, characteristic_lengthscale}
\begin{equation}
    \ket{\psi_k(t)} = \exp\Bigl(\frac{\iota}{\hbar}\epsilon_k(t)\Bigr)\ket{u_k(t)},
\end{equation}
where
\begin{equation}
\begin{aligned}
    \epsilon(k) &= \frac{1}{T}\int_0^T \dd \tau E(q_k(\tau))
                = -2\mathcal{J}_0\Bigl(\frac{A}{\hbar\omega}\Bigr)\cos{k} \\
                &= -2J_{\text{eff}}\cos{k},
\end{aligned}
\end{equation}
with the renormalized hopping strength $J_{\text{eff}} = \mathcal{J}_0(A/\hbar\omega)$ where $\mathcal{J}_0$ denotes the Bessel function of zeroth order. $\ket{u_k(t)} = \ket{u_k(t + T)}$ is a $T$-periodic Floquet state. The phase factor is tuned to unity when $A/\omega$ is chosen such that it is a root of the zeroth order Bessel function, causing the renormalized hopping parameter to vanish. This vanishing leads to the collapse of the band. As a result, an initially localized wave-packet remains localized and reproduces itself exactly after one period, hence exhibiting dynamical localization. Unless stated otherwise, we set the parameters $\hbar = 1$, $a = 1$, $J = 1$, and $\omega = 1$ for all computations.

\subsection{\label{subsec:elec_results}Results: Bloch Oscillations and Dynamical Localization}

The characteristic length-scale to observe Bloch oscillations is given in Eq. (\ref{eq:blochlength}). If $L > L_B$, Bloch oscillations can be observed with the frequency $\omega_B = aF/\hbar$, and a particle initially localized at the center of the lattice can be expected to return to the same position at every integer multiple of the Bloch period given by $T_B = 2\pi/\omega_B$. The localization of the particle can be quantified by studying the deviation of the position of the particle w.r.t an initial reference. Thus, we study the time evolution of the mean squared displacement (MSD) of a single particle
\begin{equation}
    \sigma^2(t) = \langle \hat{n}^2(t) \rangle - \langle\hat{n}(t)\rangle^2 .\label{eq:msd}
\end{equation}

In a system of size $M=7$, we consider the initial state as a wave-packet localized at the center of the lattice. While adapting this problem to a boson sampling simulation, this is expressed as a single photon initialized at the center of the lattice, with the Fock state $\ket{0001000}$. The steps taken to put together a boson sampling circuit are as discussed in Section \ref{subsec:methodology}. During the evolution, at each timestep, a new interferometer unitary is defined, and we obtain the Fock basis probabilities using a method in Strawberry Fields. Thus, we only have access to $|\psi(t)|^2$. 
With this in hand, we express the MSD in the following form
\begin{eqnarray}
    \sigma^2(t) = \sum_i(i)^2P_i(t) - \Bigl(\sum_i(i)P_i(t)\Bigr)^2,\quad 
    P_{i}(t)=|\psi_{i}|^{2}\nonumber\\
\end{eqnarray}
where $P_i(t)=|\psi_{i}(t)|^{2}$ gives the probability of finding the photon at lattice site $i$ at time $t$, and $|\psi(t)\rangle$ is the time-evolved state at time $t$, as obtained via the boson sampling simulation.

We can observe Bloch oscillations in Fig. (\ref{fig:static_heatmap}), which depicts the time-evolution of the on-site populations $P_i(t) = |\langle i\ket{\psi(t)}|^2$ at each site. In Fig. (\ref{fig:msd_bloch}) we compare results of time evolution of the initial state according to the Hamiltonian (\ref{hamiltonian_bloch}) of system size $M=7$, for a single particle. In Fig. \ref{fig:msd_bloch} (a), at field strength $F = 1$, we observe dynamics that are not strictly periodic, owing to boundary effects that arise when the system size is smaller than the localization length i.e. Bloch length $L_B$. On the other hand, in Fig. \ref{fig:msd_bloch} (b), for a field strength of $F = 2$, the Bloch length becomes less than the system size, and the particle can be seen to be localized within the vicinity of its original position exhibiting perfectly periodic behaviour. This periodic decline of the MSD to zero clearly indicates the prevalence of Bloch oscillations. The Fock backend of Strawberry Fields currently supports a maximum of $23$ modes for a photonic circuit. Due to this limitation, in order to clearly observe Bloch oscillations, it is necessary to adequately select the strength of the electric field such that the localization length $L_B$ is less than the system size.

Now, in the case of a sinusoidal driving, the mean squared displacement of an initially localized wave-packet is unbounded for an arbitrary ratio of $A/\omega$, which can be seen in the time-evolution of onsite populations computed via boson sampling and numerics in Fig. (\ref{fig:dl}) (a). However, when $A/\omega$ is a root of the Bessel function of zeroth order\cite{characteristic_lengthscale}, the renormalized hopping strength $\mathcal{J}_{0}(A/\hbar\omega)$ tends to vanish, and the dynamics is bounded and periodic (Fig. \ref{fig:dl} (a), (b)). Hence, an initially localized wave-packet returns to the initial state after a certain time-period. Fig. \ref{fig:dl} (b) illustrates a comparison of the MSD as obtained via numerics and boson sampling. Again, periodic fall of the MSD to zero is a signature of localization.
It can be seen in both cases (Bloch oscillations and dynamical localization) that as the amplitude of the drive is increased, the propagation of the particle gets more suppressed, and the particle becomes more localized.

We can see an excellent agreement in all the results recovered from boson sampling in comparison to numerics.
\section{\label{sec:aah}Aubry-Andr\'e-Harper Model}
The one-dimensional Aubry-Andr\'e-Harper (AAH) model, governed by a quasiperiodic potential hosts a delocalization-localization phase transition even in one dimension\cite{Harper1955, aubry_1980_analyticity, phasetrans_nonhermitian_aah, asharma_aah_delocalizedeig, entropy_otoc_aah, aah_spectral_corr, aah_optical_superlattices, aahproperties_pwave}. The model Hamiltonian is defined as
\begin{equation}
    \hat{H} = -J\sum_{i=1}^{N}(\hat{c}_i^\dag \hat{c}_{i+1} + \text{h.c.}) + \sum_{i=1}^N\lambda\cos{(2\pi\alpha i + \delta)}\hat{c}_i^\dag\hat{c}_i,\label{aah_hamiltonian}
\end{equation}
where $J$ is the strength of the hopping, $\lambda$ is the strength of the quasiperiodic potential, and $\delta$ is an arbitrary phase. It is known that when the $\alpha$ is chosen to be the inverse of the golden mean i.e., $(\sqrt{5} - 1)/2$, all eigenstates of a single particle are in the extended/delocalized phase for $\lambda<2J$, multifractal at the critical point with $\lambda = 2J$ and localized for $\lambda>2J$~\cite{andersontransitions} . Following the standard convention, we set $\alpha$ to the inverse of the golden mean. The model is also self-dual\cite{aubry_1980_analyticity, selfdual}, hence at the critical point, the AAH model in position space maps to itself in momentum space. All the data presented in the following section has been averaged over $50$ realizations of the arbitrary phase $\delta$, with parameters $J=1$, $\alpha = (\sqrt{5}-1)/2$, $\hbar=1$.

\subsection{\label{subsec:aah_results}Results: Phase Transition in AAH Model}

In this section, we analyse AAH model using boson sampling circuits, and present the dynamics of root-mean-squared displacement, onsite populations, second-order dynamical participation entropy, and inverse participation ratio to capture the delocalization to localization phase transition on increasing the quasiperiodic disorder strength in AAH model. Again, we follow the steps highlighted in Secion \ref{subsec:methodology} for simulating these dynamics using boson sampling.

Scaling of participation entropies with system size is a useful measure to study signatures of localization\cite{aah_expobs_npjquantum}. The dynamical participation entropy characterizes how rapidly a wavefunction $\ket{\psi(t)}$ spreads over the Hilbert-space. In our case, as time-evolution causes the wavefunction to spread in the Fock space, the participation entropy increases with time up to some saturation point.
\begin{figure}[ht]
    \centering
        \includegraphics[width=0.95\linewidth]{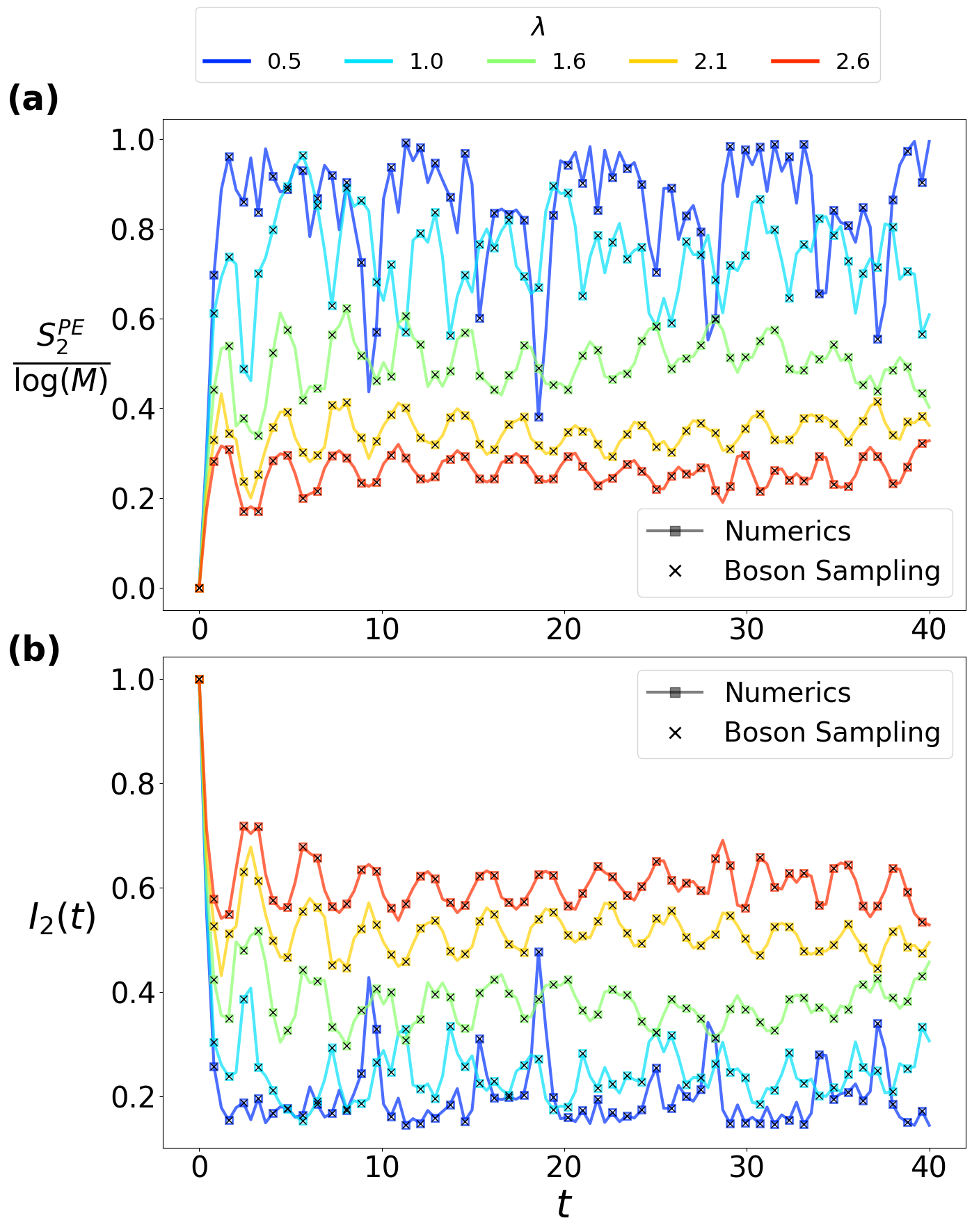}
        \caption{Dynamics of AAH model for varying values of quasi-periodic disorder strength $\lambda$. \textbf{(a)} Dynamics of second-order participation entropy $S_2^{PE}(t)$, normalized by the natural logarithm of the system size ($M=7$). \textbf{(b)} Dynamics of Inverse Participation Ratio $I_2$(t). The data for each plot was computed for evolution upto $t = 40$ discretized into 100 timesteps, and averaged over 50 realizations of $\delta$.}
    \label{fig:aah_other}
\end{figure}
\begin{figure*}[ht]
    \includegraphics[width=0.44\linewidth]{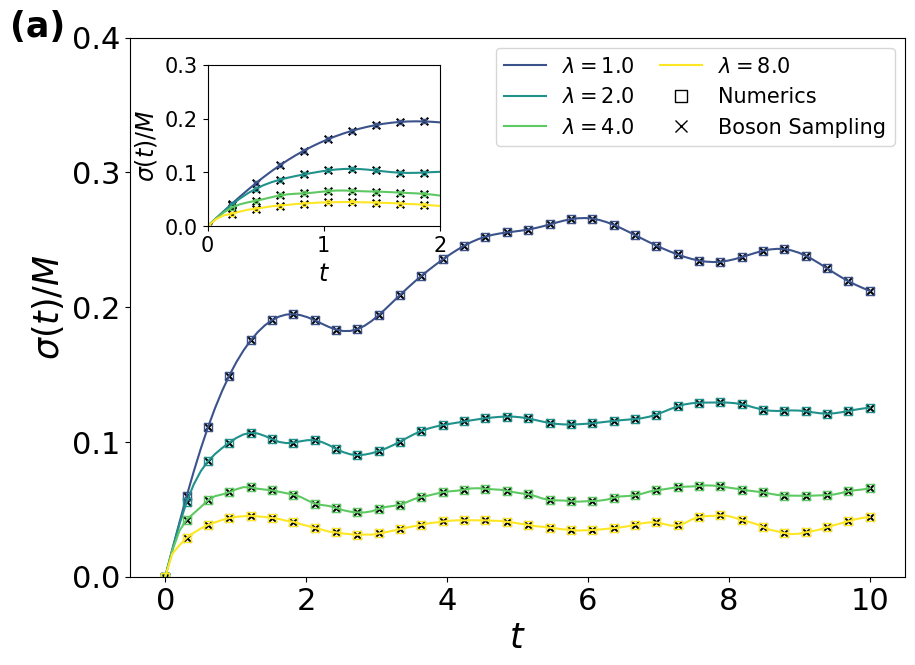}
    \includegraphics[width=0.44\linewidth]{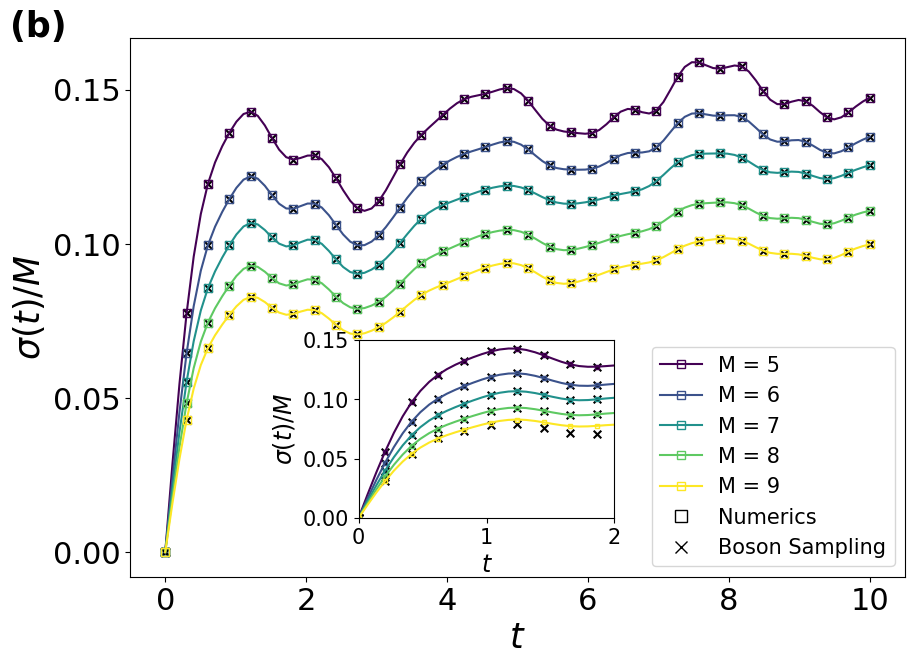}
    \caption{\textbf{(a)} Dynamics of root mean squared displacement $\sigma(t)$ of a single particle, normalized by system size ($M=7$) (Eq. \ref{eq:msd}). The behaviour in the delocalized phase at a smaller timescale is almost linear, indicating diffusive transport. As $\lambda$ increases, the transport of the particle is suppressed and the particle gets more localized. \textbf{(b)} Root mean squared displacement (at $\lambda=2$) of the particle as a function of time for varying system sizes. This figure highlights the influence of boundary effects when smaller systems are being considered, indicating the system size to be considered must be larger than the characteristic-length scale of the displacement of the particle. The data for each system size is averaged over 50 values of the arbitrary phase parameter $\delta$, with evolution upto $t=10$ (100 timesteps) and $t=2$ (30 timesteps). Every third datapoint has been marked, with the square markers denoting numerical results and the cross markers denoting boson sampling results.}
    \label{fig:rmsd}
\end{figure*}

The $q^{th}$ order dynamical participation entropy, also known as the R\'enyi entropy, is defined as\cite{s2pe_ref2, multifractality, s2pe_ref1, aah_expobs_npjquantum}.
\begin{equation}
    S_q^{PE}(t) = \frac{1}{1-q}\log{\Bigl(\sum_i^MP_i(t)^q\Bigr)},\label{qth_order_P}
\end{equation}
where $P_i(t)$ is the time-evolved onsite population at site $i$. The $q \to 1$ case recovers the well known Shannon entropy\cite{multifractality}. For a configuration space of size $\mathcal{N}$ (in our case, $\mathcal{N} = {{M+N-1} \choose N}$, in the thermodynamic limit $\mathcal{N} \to \infty$, the state is (a) localized if $S_q$ is bounded by a constant, (b) delocalized if $S_q/\log{\mathcal{N}} \to 1$ (c) multifractal if $S_q/\log{\mathcal{N}} \to D_q$, where $D_q$ denotes the fractal dimension\cite{multifractality, multifractality2, multifractality3} of the wavefunction. We study $S_2^{PE}$ as a function of time

\begin{equation}
    S_2^{PE}(t) = -\log{\Bigl(\sum_i^MP_i(t)^2\Bigr)}.\label{s2pe}
\end{equation}

Localization properties can also be quantified by the inverse participation ratio (IPR)\cite{ipr1, andersontransitions}, defined as

\begin{equation}
    I_2(t) = \sum_{i}^{M}P_i(t)^2.
\end{equation}
IPR helps to quantify the spread of the wavefunction over all the energy eigenstates. This is especially useful for wavefunctions in superpositions over multiple states. For normalized wavefunctions in an extended state, $I_2 \sim 1/\sqrt{\mathcal{N}}$, whereas for a strongly localized state, $I_2 \sim 1$, with intermediate behaviour in the critical phase\cite{ipr1, ipr2, aah1, loc_persisting_aperiodic, aah_expobs_npjquantum, weld2024}. 

In Fig. \ref{fig:aah_other} (a), we illustrate the dynamics of the second order participation entropy $S_2^{PE}$, upto late times. We can see that after a fast initial relaxation, the participation entropy in all cases shows an oscillatory behaviour around a particular value. For smaller values of the quasiperiodic disorder strength ($\lambda < 2$), we can see that $S_2^{PE}/\log{\mathcal{N}}$ tends to oscillate close to 1, denoting the delocalized phase. As $\lambda$ increases, transport is suppressed and $S_2^{PE}$ at late times does continues to oscillate around a much lower value and does not exceed some constant upper limit -- this constant value depends on the magnitude of $\lambda$, and is independent of the size of the configuration space. This is a signature of the system being in the localized phase.

In Fig. \ref{fig:aah_other} (b), we show the late-time dynamics of the inverse participation ratio $I_2$. A low IPR value suggests that the wavefunction is distributed over a large number of sites. IPR remains low for low values of $\lambda$ before the critical point, showing delocalization characteristic of the extended phase. IPR can be seen to increase and saturate at a higher value as $\lambda$ crosses the critical point, going into the localized phase. 

The gradual decrease in the root-mean-squared displacement (RMSD) of a single particle (normalized by the system size) is a good tool to characterize the phase transition using dynamics\cite{rmsd1}. The effect of varying $\lambda$ values and system sizes at different timescales is illustrated in Fig. \ref{fig:rmsd} (a) and (b) respectively. The root-mean-squared displacement as a function of time is derived by taking the square root of Eq. (\ref{eq:msd}) and scaling according to the system size $M$. In Fig. \ref{fig:rmsd} (a), we observe the dynamics of root-mean-squared displacement in an AAH system of size $M=7$, for four different values of $\lambda$, at two different timescales. It can be seen that as $\lambda$ increases and the system localizes, the RMSD gets bounded by a constant of lower magnitude, and continues oscillating after an initial saturation. At a timescale before this saturation, we can see that for $\lambda<2$, the behaviour of RMSD is almost linear w.r.t time, indicating diffusive transport. Fig. \ref{fig:rmsd} (b) highlights how the RMSD varies at different system sizes. The finite size of the lattices results in the boundaries having a notable effect on the transport of the particle. A higher RMSD at lower system sizes can be attributed to the influence of boundary effects like reflections in cases wherein the characteristic length-scale of the dynamics is larger than the lattice size.
\begin{figure}
        \centering
        \includegraphics[width=\linewidth]{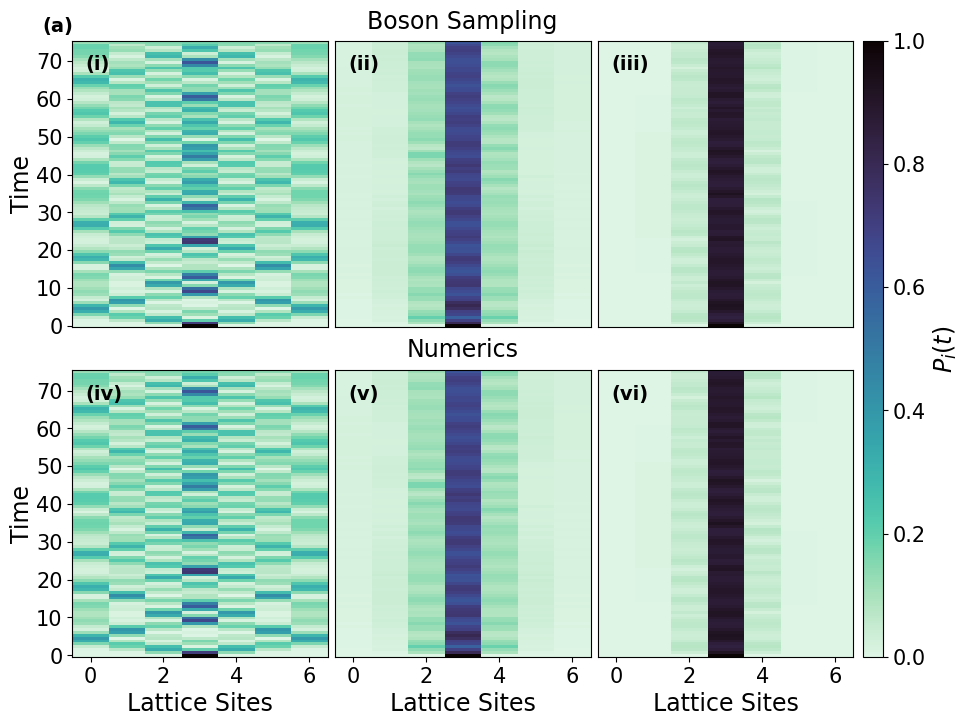}
    \caption{Evolution of onsite populations $P_i(t)$ with time for AAH model. \textbf{(a)} Late-time evolution of on-site populations for three values of $\lambda$: (i, iv) $\lambda=0.5$, (ii, v) $\lambda=2$, and (iii, vi) $\lambda=5$. The initial state in each case is $\ket{0001000}$. \textbf{(i), (ii), (iii)} respectively show the dynamics obtained using boson sampling in the extended/delocalized phase, critical phase and localized phase. 
 \textbf{(iv), (v), (vi)} show a corresponding numerical comparison for each of the three cases listed above. All cases consider the parameters $J=1$, time $t=75$ discretized over 100 timesteps. The data is averaged over 50 realizations of the arbitrary phase $\delta$, equally spaced in the interval $[-\pi, \pi]$.}
    \label{fig:aah_heatmap}
\end{figure}

We display the time-evolution of on-site populations in the Aubry-Andr\'e-Harper model in Fig. (\ref{fig:aah_heatmap}). It can be seen how an increase in $\lambda$ blocks the transport of the single-particle wave-packet. At $\lambda = 2J = 2$, we observe the critical phase where the transport is not completely suppressed, however the particle only tends to oscillate in the close vicinity of its initial position, with the population at the initial position remaining higher than 0.5. For systems with $\lambda > 2$, the particle becomes fully localized, with minimal probability of finding it anywhere other than its original position at the start of the dynamics.

We have therefore probed Bloch oscillations, dynamical localization and the phase transition in the one-dimensional Aubry-Andr\'{e}-Harper model by studying the dynamics using boson sampling, and confirmed that the results obtained closely agree with numerics.

\section{\label{discussion}Summary and Conclusion}
\begin{figure*}
    \centering
    \includegraphics[width=\textwidth]{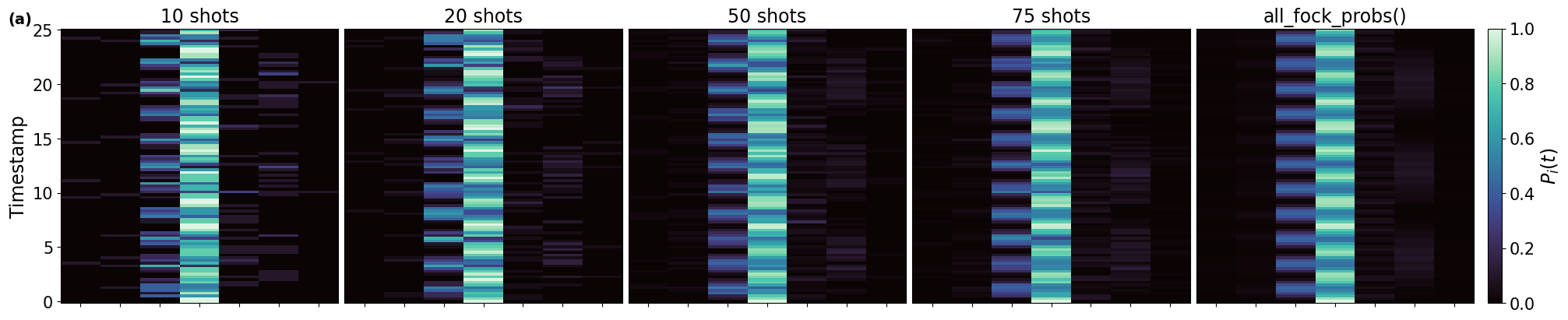}
    \includegraphics[width=\textwidth]{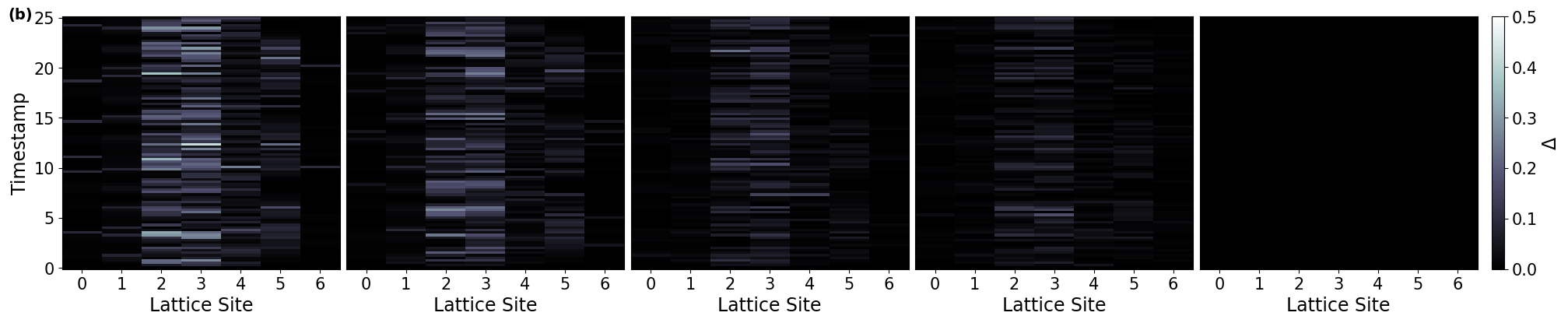}
    \cprotect\caption{\textbf{(a)} Comparison of time evolution of onsite populations of a single particle in the critical phase $(\lambda=2)$ of the AAH model. We can see increasingly accurate behaviour with an increase in the number of shots. The last column depicts results obtained using the \verb|all_fock_probs()| method, which is increasingly accurate to the numerical results. \textbf{(b)} A corresponding comparison of the error parameter $\Delta$ across different numbers of shots. We can see that a higher number of shots reduces the error, thereby giving more accurate results. All plots consider an initial Fock state $\ket{0001000}$, evolving for time $t=25$ discretized into 100 timesteps, with the parameters $J = 1, \delta=0, M = 7, \text{ and } \lambda=2$.}
    \label{fig:shots_analysis}
\end{figure*}
In this work, we devised an algorithm to obtain the dynamics of quantum systems using boson sampling. We have observed the effects of static and time-periodic electric field on the transport of a single particle in a nearest-neighbour tight-binding chain. Furthermore, we observe localization properties of Bloch oscillations and dynamical localization using mean-squared displacement of a single particle and evolution of onsite populations. We also characterize the phase transition from extended to localized phase in the AAH model with the dynamical study of participation entropies and inverse participation ratio (IPR).

From this study, we can safely conclude that boson sampling (and photonic circuits in general) can act as synthetic quantum platforms performing an analog simulation, and can therefore be used to probe dynamics. Bloch oscillations and dynamical localization can be further explored using boson sampling by finding novel ways to compute localization length, and by extending the study to transport phenomena in more complex systems such as disorded potentials. The same idea of quantum simulation can be extended to more general photonic circuits, which can encompass systems that are interacting and/or chaotic. Such simulations are possible using the same Strawberry Fields library, which has support for both Kerr and cross-Kerr interactions\cite{sfpaper, sfpaper2}. Simulations on photonic platforms can be performed at room temperature, which is especially useful, considering contemporary platforms (for example, superconducting qubits) operate at very low temperatures and therefore need extensive cooling\cite{superconductingprocessor1, superconductingprocessor2}. 

It is also possible to study spin systems on such continuous-variable platforms\cite{spinhamiltoniansoncv}. Realizing viable quantum simulation algorithms on a Gaussian boson sampling platform is another interesting open direction, which can help further simplify experimental realizations, due to the relative ease of generating Gaussian states of light as compared to single-photon states.

So far we have studied only single-particle dynamics using this method. It will be especially interesting to extend to many-body dynamics and probe whether there is any computational advantage to using boson sampling for solving many-body problems, which are known to be notoriously hard to compute classically. It should be noted, however, that using this platform for simulations of dynamics at large times will inevitably cause a rise in the circuit depth. This demands a detailed analysis of the scaling of the circuit depth with the problem size, in order to properly assess the scalability of the platform. For a single particle, boson sampling is able to replicate dynamics of an electron (which is a fermion), as we do not need to account for how it behaves in the presence of other particles based on its statistics. It has not yet been studied whether there exists an encoding/map in order to harness a photonic platform to study fermionic many-body dynamics.

As the Fock backend of the Strawberry Fields\cite{sfpaper} simulator is built on NumPy\cite{numpy}, the computations it makes are inherently classical. Hence, no comment can be made yet about whether boson sampling will be faster than classical computation for the current use cases. This presents another open problem - estimating the time required per shot measurement and overall duration of the simulation on a photonic circuit.

\section*{Acknowledgments}
We are grateful to the High Performance Computing (HPC) facility at IISER Bhopal, where large-scale calculations in this project were run. A.K acknowledges funding support for Chanakya - PG fellowship from the National Mission on Interdisciplinary Cyber Physical Systems, of
the Department of Science and Technology, Govt. of India through the I-HUB Quantum Technology
Foundation. V.T is grateful to DST-INSPIRE for her PhD fellowship. A.S acknowledges financial support from SERB via the grant (File Number:
CRG/2019/003447), and from DST via the DST-INSPIRE Faculty Award [DST/INSPIRE/04/2014/002461]. This research made use of Strawberry Fields version 0.23.0, an open-source software framework for photonic quantum computing. 

\appendix*
\section{\label{appendix:samples}On the effect of sample count}

The boson sampling simulations in this work were performed on a local state-vector simulator using the Fock backend of the Strawberry Fields library. This backend simulates quantum optical circuits in a truncated Fock basis using NumPy\cite{sfpaper, numpy}. A class called \verb|strawberyyfields.backends.BaseFockState| is used to represent quantum states in the Fock basis. There exists a method belonging to this class called \verb|all_fock_probs()| that returns the probabilities of all possible Fock basis states for the current circuit state - which will be stored in the result object upon running the simulation. This method was used to obtain results that were then compared against numerical data.\\

It should be noted that in an experimental scenario, we will have to perform multiple ``shots" of the same circuit each with a photon-counting measurement following the evolution. Performing an average over the samples thus obtained will be necessary to approximate the Fock space probabilities. A higher sample count therefore leads to greater accuracy of the result. Let us consider an example to illustrate this effect. Consider the time evolution of a photon in the critical phase of the Aubry-Andr\'e-Harper model. Fig. (\ref{fig:shots_analysis}) depicts a comparison between five instances of the same circuit, with the first four columns showing results for an increasing number of shots, while the fifth column showing results via the \verb|all_fock_probs()| method. It can be seen in Figure \ref{fig:shots_analysis} (a) that the trend gradually nears the expected behaviour as the number of shots is increased.

\begin{figure*}
    \centering
    \includegraphics[width=\linewidth]{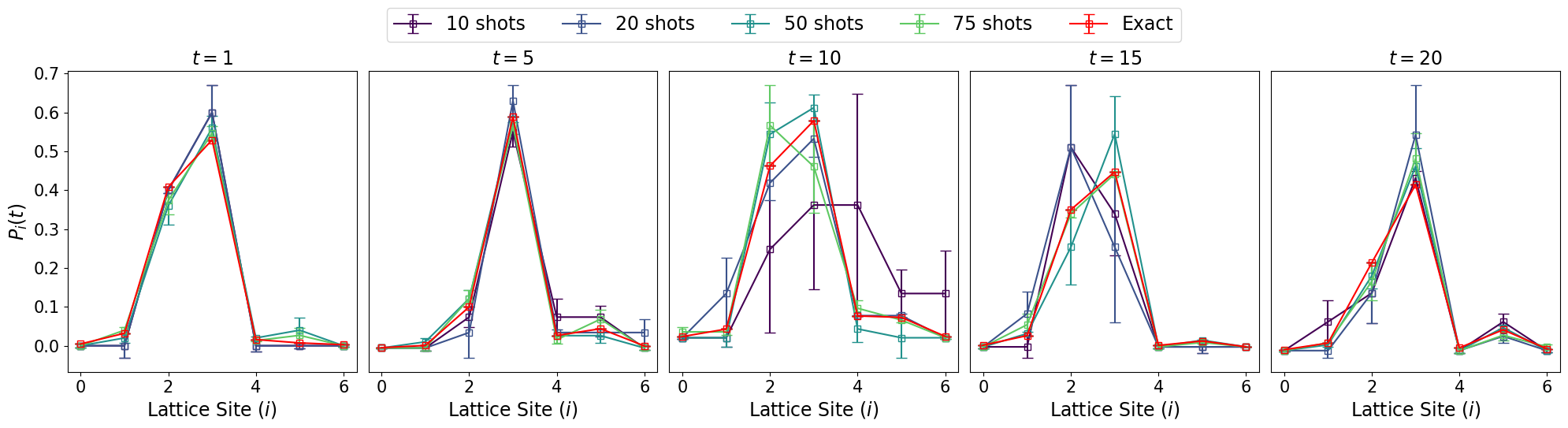}
    \cprotect\caption{Comparison of probability density of the wave-function at five representative timestamps, for varying shot counts. We consider the exact result as the one obtained from the \verb|all_fock_probs()| method. The system is an AAH model, with state at $t=0$ initialized to $\ket{0001000}$, and system parameters $J=1, \delta=0, M=7, \lambda=2$. The probability estimation at 10 shots per timestep is poor, however we can see that as the number of shots increases, the error margin $\Delta$ reduces significantly.}
    \label{fig:shots_errorbars}
\end{figure*}

We define a true error parameter $\Delta$ as the absoute difference between a probability obtained by boson sampling and the corresponding numerical result.
\begin{equation}
    \Delta(t) = |P_{\mathrm{Num}}(t) - P_{\mathrm{BS}}(t)|.
\end{equation}
We can see in Fig. \ref{fig:shots_analysis} (b), that as the number of shots increases, $\Delta$ becomes lesser, which implies a higher accuracy of the simulated result. This decrease in the error margin is also apparent in the probability density of the state. We show in Fig. (\ref{fig:shots_errorbars}) a comparison of this across multiple shot values. One may also use other metrics such as state-fidelity to study the accuracy in such a case.

\bibliography{citations}

\end{document}